\documentclass[amsmath,aps,showpacs,a4paper,10pt,superscriptaddress]{revtex4}
 \usepackage{epsf}
 \usepackage{graphicx}    % Include figure files

\begin{document}

 \newcommand{\be}[1]{\begin{equation}\label{#1}}
 \newcommand{\ee}{\end{equation}}
 \newcommand{\bea}{\begin{eqnarray}}
 \newcommand{\eea}{\end{eqnarray}}
 \def\disp{\displaystyle}

 \def\gsim{ \lower .75ex \hbox{$\sim$} \llap{\raise .27ex \hbox{$>$}} }
 \def\lsim{ \lower .75ex \hbox{$\sim$} \llap{\raise .27ex \hbox{$<$}} }

 \begin{titlepage}

 \begin{flushright}
 arXiv:1411.6218
 \end{flushright}

 \title{\Large \bf Age Problem
 in Lema\^{\i}tre-Tolman-Bondi Void Models}

 \author{Xiao-Peng~Yan\,}
 \email[\,email address:\ ]{764644314@qq.com}
 \affiliation{School of Physics, Beijing Institute
 of Technology, Beijing 100081, China}

 \author{De-Zi~Liu\,}
 \affiliation{Department of Astronomy, Peking University,
 Beijing 100871, China}

 \author{Hao~Wei\,}
 \thanks{\,Corresponding author}
 \email[\,email address:\ ]{haowei@bit.edu.cn}
 \affiliation{School of Physics, Beijing Institute
 of Technology, Beijing 100081, China}

 \begin{abstract}\vspace{1cm}
 \centerline{\bf ABSTRACT}\vspace{2mm}
 As is well known, one can explain the current cosmic acceleration by
 considering an inhomogeneous and/or anisotropic universe
 (which violates the cosmological principle), without invoking dark
 energy or modified gravity. The well-known one of this kind of
 models is the so-called Lema\^{\i}tre-Tolman-Bondi (LTB) void
 model, in which the universe is spherically symmetric and radially
 inhomogeneous, and we are living in a locally underdense void
 centered nearby our location. In the present work, we test various
 LTB void models with some old high redshift objects (OHROs).
 Obviously, the universe cannot be younger than its constituents. We
 find that an unusually large $r_0$ (characterizing the size of the
 void) is required to accommodate these OHROs in LTB void models.
 There is a serious tension between this unusually large $r_0$
 and the much smaller $r_0$ inferred from other observations
 (e.g. SNIa, CMB and so on). However, if we instead consider
 the lowest limit 1.7\,Gyr for the quasar APM 08279+5255 at
 redshift $z=3.91$, this tension could be greatly alleviated.
 \end{abstract}


 \maketitle

 \end{titlepage}

 \renewcommand{\baselinestretch}{1.0}

%============================= section 1 ===================================

\section{Introduction}\label{sec1}

Since the discovery of the current accelerated expansion of the
 universe \cite{1999ApJ...517..565P, Riess:1998cb,
 Spergel:2003cb, Hinshaw:2012aka, Tegmark:2003ud,
 Tegmark:2006az}, various models have been proposed to explain
 this mysterious phenomenon. As is well known, the modern cosmology
 is based on general relativity and the cosmological principle.
 The well-known Einstein field equations read
 $$G_{\mu\nu}=8\pi G T_{\mu\nu}\,,$$
 where $G_{\mu\nu}$ and $T_{\mu\nu}$ are the Einstein tensor
 and the stress-energy tensor respectively, and we set the
 speed of light $c=1$ throughout this work. According to the
 pillars of modern cosmology, these theoretical models can be
 categorized into the following three major types.

The first one is to modify the right hand side of Einstein
 field equations. That is, one can introduce an exotic energy
 component, namely dark energy with negative pressure
 \cite{Padmanabhan:2004av,Copeland:2006wr,Kamionkowski:2007wv},
 while general relativity still holds. The simplest candidate
 of dark energy is a tiny cosmological constant
 \cite{Nobbenhuis:2004wn, Sahni:1999gb} introduced by Einstein
 himself in 1917. As is well known, it seriously suffers from
 the fine-turning problem and the cosmological coincidence
 problem \cite{Peebles:2002gy, Sahni:1999gb,Albrecht:2006um,
 Carroll:2003qq}. To alleviate these problems, various
 dynamical models of dark energy were proposed, such as
 quintessence \cite{Caldwell:1997ii,Zlatev:1998yg,
 Steinhardt:1999nw}, phantom \cite{Caldwell:2003vq,Caldwell:1999ew},
 $k$-essence \cite{ArmendarizPicon:2000dh,
 ArmendarizPicon:2000ah,Chiba:1999ka}, quintom
 \cite{Feng:2004ad}, Chaplygin gas \cite{Kamenshchik:2001cp,
 Bento:2002ps}, vector-like dark energy \cite{ArmendarizPicon:2004pm,
 Wei:2006tn,Wei:2006gv}, holographic dark
 energy \cite{Li:2004rb}, (new) agegraphic dark
 energy \cite{Cai:2007us,Wei:2007ty,Wei:2007xu},
 hessence \cite{Wei:2005nw,Wei:2005fq}, spinor dark energy
 \cite{Boehmer:2009aw,Boehmer:2010ma,Wei:2010ad}, and so on.

The second one is to modify the left hand side of Einstein
 field equations, namely to modify general relativity on
 cosmological scale. Einstein's general relativity is checked
 to hold in the range from large scales like the solar system
 to small scales in the order of millimeter. However, there is
 no {\it a priori} reason to believe that general relativity
 cannot be modified on cosmological scales. In the literature,
 various modified gravity theories were proposed to account for
 the cosmic acceleration, for instance, $f(R)$ theory
 \cite{DeFelice:2010aj,Sotiriou:2008rp,Clifton:2011jh},
 scalar-tensor theory \cite{Clifton:2011jh,SaezGomez:2008ex},
 Dvali-Gabadadze-Porrati (DGP) model \cite{Dvali:2000hr,
 Deffayet:2000uy,Deffayet:2001pu}, Galileon
 gravity \cite{Nicolis:2008in,DeFelice:2010nf,DeFelice:2011bh},
 Gauss-Bonnet gravity \cite{Koivisto:2006ai, Nojiri:2005vv},
 $f(T)$ theory \cite{Bengochea:2008gz,Linder:2010py}, massive
 gravity \cite{Fierz:1939ix,deRham:2010kj,Hassan:2011hr}.

\hspace{-0.54mm}The third one is to give up the cosmological
 principle, and consider an inhomogeneous and/or anisotropic
 universe, without invoking dark energy or modified gravity.
 As a tenet, the cosmological principle is known to be partly
 satisfied on large scales. However, it has not been proven
 on cosmic scales $\gsim\,1\,{\rm Gpc}$~\cite{Caldwell:2007yu}.
 Obviously, our local universe is inhomogeneous and anisotropic
 on small scales. On the other hand, the nearby sample has
 been examined for evidence of a local ``Hubble Bubble''
 \cite{Zehavi:1998gz}. It is reasonable to imagine that we are
 living in a locally underdense void. If the cosmological
 principle is relaxed, it is possible to explain the apparent
 cosmic acceleration in terms of a peculiar distribution of
 matter centered upon our location \cite{Celerier:1999hp,
 Celerier:2000pz,Celerier:2000ew,Celerier:2007jc,
 Barrett:1999fd,Tomita:2000jj,Tomita:2001gh,Iguchi:2001sq}.
 In the literature, the cosmological principle has been tested
 by using e.g. type~Ia supernovae (SNIa) \cite{Celerier:1999hp,
 Clifton:2008hv,Marra:2010pg}, cosmic microwave background
 (CMB) \cite{Caldwell:2007yu,Hansen:2004vq,Singal:2013aga,
 Moffat:2005yx,Alnes:2006pf,Clifton:2009kx,Clarkson:2010ej,
 Moss:2010jx}, time drift of cosmological
 redshifts \cite{Uzan:2008qp,Quartin:2009xr}, baryon acoustic
 oscillations (BAO) \cite{Bolejko:2008cm,February:2012fp,
 Zibin:2008vk}, integrated Sachs-Wolfe effect \cite{Tomita:2009wz},
 galaxy surveys \cite{Labini:2010qx}, kinetic
 Sunyaev Zel'dovich effect \cite{Zhang:2010fa,
 Valkenburg:2012td,Bull:2011wi,Marra:2011ct,Moss:2011ze},
 observational $H(z)$ data \cite{Zhang:2012qr,Wang:2011kj},
 gamma-ray bursts \cite{Meszaros:2009ux}, growth of large-scale
 structure~\cite{Ishak:2013vha}, and so on. It is found that
 the violation of cosmological principle can be consistent with
 most of  these observations (in fact few observations slightly
 favor the violation of cosmological principle). Therefore, it
 is reasonable to consider an inhomogeneous and/or anisotropic
 universe. In the literature, the well-known models violating
 cosmological principle are the so-called
 Lema\^{\i}tre-Tolman-Bondi (LTB) void models
 \cite{Lemaitre:1933gd,Tolman:1934za,Bondi:1947fta}. In LTB
 void models, the universe is spherically symmetric and
 radially inhomogeneous, and we are living in a locally
 underdense void centered nearby our location. The Hubble
 diagram inferred from lines-of-sight originating at the
 center of the void might be misinterpreted to indicate cosmic
 acceleration. In fact, LTB void models can be consistent with
 (even slightly favored by) the observations mentioned above.

In the present work, we try to test LTB void models with the
 age of the universe. Obviously, the universe cannot be younger
 than its constituents. In history, the age problem played an
 important role in cosmology for many times. However, we
 should clarify the two meanings of age problem. The first
 meaning is that the total age of the universe (namely the
 age measured at present day, or, redshift $z=0$) cannot be
 smaller than the age of the oldest known objects (e.g.
 globular clusters, galaxies, quasars) in our universe.
 Historically, the matter-dominated Friedmann-Robertson-Walker
 (FRW) model without cosmological constant can be ruled out
 \cite{Alcaniz:1999kr} because its total age is smaller than
 the ages inferred from old globular clusters, unless the
 Hubble constant is extremely low or the universe is extremely
 open. In the literature, one might consider a variant of this
 type of age problem. For instance, the authors
 of~\cite{Lan:2010ky,Liu:2013jga} reconstructed LTB model from
 $\Lambda$CDM model by requiring they share the same expansion
 history (luminosity distance, light-cone mass density, angular
 diameter distance $d_A(z)$, Hubble parameter $H(z)$), and
 found that the total age of the universe inferred from LTB
 model is much smaller than the one inferred from $\Lambda$CDM
 model ($t_{\rm \Lambda CDM}-t_{\rm LTB}\sim 2.4\,{\rm Gyr}$).
 However, strictly speaking, this variant of age problem is
 not the real age problem, since LTB model is the reconstructed
 one, and the total age of the universe is not compared with
 the real age of old objects (e.g. globular clusters, galaxies,
 quasars). So, we do not consider this kind of age problem in
 the present work.

Instead, here we consider the second meaning of age problem,
 namely the age of the universe at any high redshift $z>0$
 (rather than the total age at present day, $z=0$) cannot be
 younger than its constituents at the same redshift. Obviously,
 in this case the age problem becomes more serious than the
 first one. There are some old high redshift objects (OHROs)
 considered extensively in the literature, for instance, the
 3.5\,Gyr old galaxy LBDS 53W091 at redshift
 $z=1.55$~\cite{Dunlop:1996mp,Spinrad:1997md}, the 4.0\,Gyr old
 galaxy LBDS 53W069 at redshift $z=1.43$~\cite{Dunlop:1999xx}.
 In addition, the old quasar APM 08279+5255 at redshift
 $z=3.91$~\cite{Hasinger:2002wg,Komossa:2002cn} is also
 used extensively. Its age is estimated to be
 2.0\,--\,3.0\,Gyr~\cite{Hasinger:2002wg,Komossa:2002cn}.
 In~\cite{Friaca:2005ba}, by using a different method, its age
 is reevaluated to be 2.1\,Gyr. To assure the robustness of our
 analysis, we use the most conservative lower age estimate
 2.0\,Gyr for the old quasar APM 08279+5255 at redshift
 $z=3.91$ throughout the present work. In the literature, these
 three OHROs have been extensively used to test various dark
 energy models (see e.g. \cite{Alcaniz:1999kr,Friaca:2005ba,
 darkenergytested:01,darkenergytested:02,
 Yang:2009ae,Wei:2007ig,Zhang:2007ps,Wei:2010hs,Wang:2010su})
 and modified gravity models (see
 e.g. \cite{Capozziello:2007gr,Movahed:2007cs,Movahed:2007ie,
 Movahed:2007ps,Pires:2006rd}). In the present work, we will
 use them to test various LTB void models.

The rest of this paper is organized as follows.
 In Sec.~\ref{sec2}, we briefly review the main points of LTB
 model. In Sec.~\ref{sec3}, we test various LTB void models
 with OHROs. In Sec.~\ref{sec4}, we discuss the possibility to
 alleviate the age problem. In Sec.~\ref{sec5}, we give the
 brief conclusion and discussion.

%============================= section 2 ===================================

\section{The LTB model}\label{sec2}

In the LTB void model, the universe is spherically symmetric
 and radially inhomogeneous, and we are living in
 a locally underdense void centered nearby our location. The
 dynamic of a spherically symmetric dust universe is described
 by the LTB solution to Einstein field equations. It was
 firstly proposed by Lema\^{\i}tre~\cite{Lemaitre:1933gd},
 then was further discussed by Tolman \cite{Tolman:1934za} and
 Bondi \cite{Bondi:1947fta}. The LTB metric, in comoving
 coordinates ($r$, $\theta$, $\phi$) and synchronous time $t$,
 is given by \cite{Lemaitre:1933gd,Tolman:1934za,Bondi:1947fta} (see
 also e.g. \cite{Zhang:2012qr,Wang:2011kj,GarciaBellido:2008nz,
 Lapiedra:2013nka})
 \be{eq1}
 ds^2 = -dt^2 + \frac{A^{\prime\,2}(r,t)}{1-k(r)}\,dr^2
 +A^2(r,t)\,d\Omega^2\,,
 \ee
 where $d\Omega^2=d\theta^2+\sin^2\theta\,d\phi^2$; a prime
 denotes a derivative with respect to $r$, and $k(r)$ is an
 arbitrary function of $r$, playing the role of spatial
 curvature. Note that it reduces to the well-known FRW
 metric if $A(r,t)=a(t)\,r$ and $k(r)=kr^2$.
 The stress-energy tensor of the mass source is given by
 \be{eq2}
 T^\nu_\mu=-\rho_{_M}(r,t) \,\delta^\nu_0 \,\delta^0_\mu\,,
 \ee
 where $\rho_{_M}$ is the energy density of dust matter. The
 Einstein field equations read \cite{Enqvist:2006cg,Enqvist:2007vb,
 GarciaBellido:2008nz,Wang:2011kj,Alnes:2005rw}
 \bea
 &&H_\perp^2 + 2H_\perp H_\parallel + \frac{k(r)}{A^2}+
 \frac{k^\prime(r)}{AA^\prime}=8\pi G\rho_{_M}\,,\label{eq3}\\
 &&\dot{A}^2 + 2A\ddot{A} + k(r)=0\,,\label{eq4}
 \eea
 where a dot denotes a derivative with respect to $t$, and
 \bea
 H_\perp (r,t)\equiv\frac{\dot{A}(r,t)}{A(r,t)}\,,\label{eq5}\\
 H_\parallel (r,t)
 \equiv\frac{\dot{A^\prime}(r,t)}{A^\prime(r,t)}\,,\label{eq6}
 \eea
 are the expansion rates at the transverse and longitudinal
 directions, respectively. Integrating Eq.~(\ref{eq4}), we
 obtain \cite{Celerier:2011zh,Celerier:2012xr,GarciaBellido:2008nz,
 Enqvist:2006cg,Enqvist:2007vb,Alnes:2005rw}
 \be{eq7}
 \dot{A}^2 (r,t) = \frac{2M(r)}{A(r,t)} - k(r)\,,
 \ee
 where $M(r)$ is an arbitrary function (the factor 2 is
 introduced just for convenience; one should be aware of the
 different symbol conventions in the relevant references).
 If $M(r)$ and $k(r)$ are given, one can obtain $A(r,t)$ by
 directly solving Eq.~(\ref{eq7}). For convenience, we instead
 try to find the parametric solutions for~it.
 Following e.g.~\cite{Goode:1982pg,Krasinski:1997xx,Celerier:2011zh,
 Celerier:2012xr}, we recast Eq.~(\ref{eq7}) as
 \be{eq8}
 \frac{\dot{A}^2(r,t)}{|k(r)|}
 =-\tilde{k}+\frac{2M(r)}{A(r,t)\,|k(r)|}\,,
 \ee
 to normalize $\tilde{k}\equiv k(r)/|k(r)|=+1,\,-1,\,0$ for
 $k(r)>0$, $k(r)<0$, $k(r)=0$, respectively. The solutions
 of Eq.~(\ref{eq8}) can be written implicitly in terms of an
 auxiliary variable $\eta$ as \cite{Goode:1982pg}
 \be{eq9}
 A(r,t) = \frac{M(r)}{|k(r)|}\frac{ds(\eta)}{d\eta}\,,
 ~~~~{\rm with}~~~~
 t-t_B (r)=\frac{M(r)\,s(\eta)}{|k(r)|^{3/2}}\,,
 \ee
 where $t_B (r)$ is actually a ``constant'' of integration.
 Therefore, Eq.~(\ref{eq8}) becomes an ordinary differential
 equation of the function $s(\eta)$,
 \be{eq10}
 \left[\frac{d^2 s(\eta)}{d\eta^2}\right]^2=
 -\tilde{k}\left[\frac{ds(\eta)}{d\eta}\right]^2
 +2\,\frac{ds(\eta)}{d\eta}\,,
 \ee
 whose solutions are given by \cite{Goode:1982pg}
 \be{eq11}
 s(\eta)=
 \begin{cases}
 \eta-\sin\eta   &  {\rm for}~~~ \tilde{k}=+1\,,\\[2mm]
 \sinh\eta-\eta & {\rm for}~~~ \tilde{k}=-1\,,\\[2mm]
 \eta^3/6  &  {\rm for}~~~ \tilde{k}=0\,.
 \end{cases}
 \ee
 Substituting Eq.~(\ref{eq11}) into Eq.~(\ref{eq9}), the
 parametric solutions of Eq.~(\ref{eq7}) read
 (see e.g. \cite{Wang:2011kj,GarciaBellido:2008nz,Krasinski:1997xx,
 Celerier:2011zh,Celerier:2012xr})
 \bea
 A(r,t)=\frac{M(r)}{k(r)}(1-\cosh\eta)\,,~~~
 t-t_B (r)= \frac{M(r)}{\left[-k(r)\right]^{3/2}}\,(\sinh\eta-\eta)
 ~~~~{\rm for}~~~k(r)<0\,,\label{eq12}\\[2.4mm]
 A(r,t)=\frac{M(r)}{k(r)}(1-\cos\eta)\,,~~~
 t - t_B (r) = \frac{M(r)}{\left[k(r)\right]^{3/2}}\,(\eta-\sin\eta)
 ~~~~{\rm for}~~~k(r)>0\,,\label{eq13}\\[2mm]
 A(r,t)=\left[\frac{9M(r)}{2}\right]^{1/3}
 \left[t-t_B(r)\right]^{2/3}
 ~~~~{\rm for}~~~k(r)=0\,,\, \label{eq14}
 \eea
 where ${t}_B(r)$ is an arbitrary function of $r$, usually
 interpreted as the ``bang time'' due to singularity behavior
 at $t=t_B$. Substituting Eq.~(\ref{eq7}) into Eq.~(\ref{eq3}),
 we have \cite{Celerier:2011zh,Celerier:2012xr,GarciaBellido:2008nz,
 Enqvist:2006cg,Enqvist:2007vb,Alnes:2005rw}
 \be{eq15}
 \frac{2M^\prime (r)}{A^\prime A^2} = 8\pi G \rho_{_M}\,.
 \ee
 Considering Eq.~(\ref{eq7}) at the present day ($t=t_0$),
 it can be recast as
 \be{eq16}
 1=\frac{2M(r)}{H_{\perp 0}^2 (r) A_0(r)^3}
 -\frac{k(r)}{H_{\perp 0}^2 (r) A_0(r)^2}
 \equiv\Omega_M(r)+\Omega_K(r)\,,
 \ee
 where the subscript ``0'' indicates the present value
 of corresponding quantity, i.e., $A_0 (r)=A(r,t=t_0)$,
 $H_{\perp 0}(r)=H_{\perp}(r,t=t_0)$. Therefore, we can
 parameterize the functions $M(r)$ and $k(r)$ as
 \cite{GarciaBellido:2008nz,Enqvist:2006cg,Enqvist:2007vb}
 \bea
 && 2M(r)=H^2_{\perp 0}(r)\,\Omega_M(r)\, A_0^3 (r)\,,
 \label{eq17}\\[1mm]
 && -k(r)=H^2_{\perp 0}(r)\,\Omega_K(r)\, A_0^2 (r)\,, \label{eq18}
 \eea
 where $\Omega_K(r)=1-\Omega_M(r)$. Noting Eq.~(\ref{eq15}),
 it is easy to see that $\Omega_M$ and $\Omega_K$ defined in
 Eqs.~(\ref{eq17}) and (\ref{eq18}) can reduce to the present
 fractional densities of FRW cosmology if $A(r,t)=a(t)\,r$ and
 $k(r)=kr^2$ while $H_{\perp 0}$ and $\Omega_M$ are spatially
 homogeneous. So, the above parameterizations are justified.
 Substituting Eqs.~(\ref{eq17}), (\ref{eq18})
 into Eqs.~(\ref{eq12})---(\ref{eq14}), we obtain the total
 cosmic age as a function of $r$ \cite{Wang:2011kj}, namely
 \be{eq19}
 t_0-t_B(r) = \frac{\mathcal{F}(\Omega_M)}{H_{\perp0}(r)}\,,
 \ee
 in which the function ${\cal F}(x)$ is defined by
 \be{eq20}
 \mathcal{F}(x)\equiv
 \begin{cases}
 \disp\frac{-\sqrt{x-1}
 +x\sin^{-1} \sqrt{\frac{x-1}{x}}} {\left(x-1\right)^{3/2}}
 & ~~{\rm for}~~~ x>1\,, \\[5mm]
 2/3 & ~~{\rm for}~~~ x=1\,, \\[1mm]
 \disp\frac{\sqrt{1-x}
 -x\sinh^{-1}\sqrt{\frac{1-x}{x}}}{(1-x)^{3/2}}
 & ~~{\rm for}~~~ x<1\,.
 \end{cases}
 \ee
 Furthermore, to compare our theoretical models
 with observations, we need to associate the coordinates with
 redshift $z$. For an observer located at the center $r=0$,
 by symmetry, incoming light travels along radial null
 geodesics, $ds^2=d\Omega^2=0$, and hence
 we have \cite{GarciaBellido:2008nz}
 \be{eq21}
 \frac{dt}{dr}=-\frac{A^\prime (r,t)}{\sqrt{1-k(r)}}\,,
 \ee
 where the minus sign is due to $dt/dr<0$, namely time
 decreases when going away. Together with the redshift
 equation \cite{Enqvist:2006cg,Enqvist:2007vb,
 GarciaBellido:2008nz,Alnes:2005rw}
 \be{eq22}
 \frac{d\ln(1+z)}{dr}=\frac{\dot{A}^\prime (r,t)}{\sqrt{1-k(r)}}\,,
 \ee
 we can write a parametric set of differential equations
 \cite{GarciaBellido:2008nz}
 \bea
 &&\frac{dt}{d\ln(1+z)}
 =-\frac{A^\prime (r,t)}{\dot{A}^\prime (r,t)}\,,\label{eq23}\\[1mm]
 &&\frac{dr}{d\ln(1+z)}
 =\frac{\sqrt{1-k(r)}}{\dot{A}^\prime (r,t)}\,.\label{eq24}
 \eea
 Once the functions $\Omega_M(r)$ and $H_{\perp0}(r)$ characterizing
 LTB model are given, substituting Eqs.~(\ref{eq17}) and
 (\ref{eq18}) into Eq.~(\ref{eq7}), the scale function $A(r,t)$ can
 be found by solving the resulting differential equation. Then, one
 can obtain $t(z)$ and $r(z)$ as functions of redshift $z$
 from Eqs.~(\ref{eq23}) and (\ref{eq24}) with the initial conditions
 $r(z=0)=0$ and $t(z=0)=t_0$. Note that in solving Eq.~(\ref{eq7}),
 the parametric solutions given in Eqs.~(\ref{eq12})---(\ref{eq14})
 are useful. One can do this numerically using a modified version of
 the code easyLTB~\cite{GarciaBellido:2008nz} (see e.g.
 \cite{Wang:2011kj} for a brief technical illustration; however, one
 should be careful of the typos in \cite{Wang:2011kj}, and the
 different symbol conventions in the relevant references, e.g.
 \cite{Wang:2011kj,Enqvist:2006cg,Enqvist:2007vb,
 GarciaBellido:2008nz,Krasinski:1997xx,Celerier:2011zh,
 Celerier:2012xr,Alnes:2005rw}, as well as the difference between the
 relevant references and the code easyLTB
 \cite{GarciaBellido:2008nz}). It is worth noting that the
 present scale function $A_0 (r)=A(r,t=t_0)$ of LTB model can
 be chosen to be any smooth and invertible positive function.
 Following \cite{Enqvist:2006cg,Enqvist:2007vb,GarciaBellido:2008nz,
 Wang:2011kj}, we choose the conventional gauge
 $A_0 (r)=A(r,t=t_0)=r$, which actually corresponds to set the
 present scale factor $a_0=a(t=t_0)=1$ in FRW cosmology.

%============================= section 3 ===================================

\section{Testing various LTB void models with OHROs}\label{sec3}

In the LTB void models, we are living at a special space point,
 which is close to the center of a large local underdense
 region of the universe \cite{Lemaitre:1933gd,Tolman:1934za,
 Bondi:1947fta,Alnes:2005rw,Celerier:2009sv,Vanderveld:2006rb}.
 At very large distances from the observer, the inhomogeneous
 LTB region goes to an external FRW space. Obviously, it
 violates the Copernican principle that states we do not occupy
 any special place in the universe. In the literature, it is
 found that the LTB void models can be consistent with (even
 slightly favored by) various observations mentioned in
 Sec.~\ref{sec1}. Here, we try to test various LTB void models with
 three OHROs mentioned in Sec.~\ref{sec1}, namely the 3.5\,Gyr
 old galaxy LBDS 53W091 at redshift
 $z=1.55$~\cite{Dunlop:1996mp,Spinrad:1997md}, the 4.0\,Gyr old
 galaxy LBDS 53W069 at redshift $z=1.43$~\cite{Dunlop:1999xx},
 and the 2.0\,Gyr old quasar APM 08279+5255 at redshift
 $z=3.91$~\cite{Hasinger:2002wg,Komossa:2002cn}.

%============================= Fig. Gauss 3d =================================

 \begin{center}
 \begin{figure}[tb]
 \centering
 \includegraphics[width=1.0\textwidth]{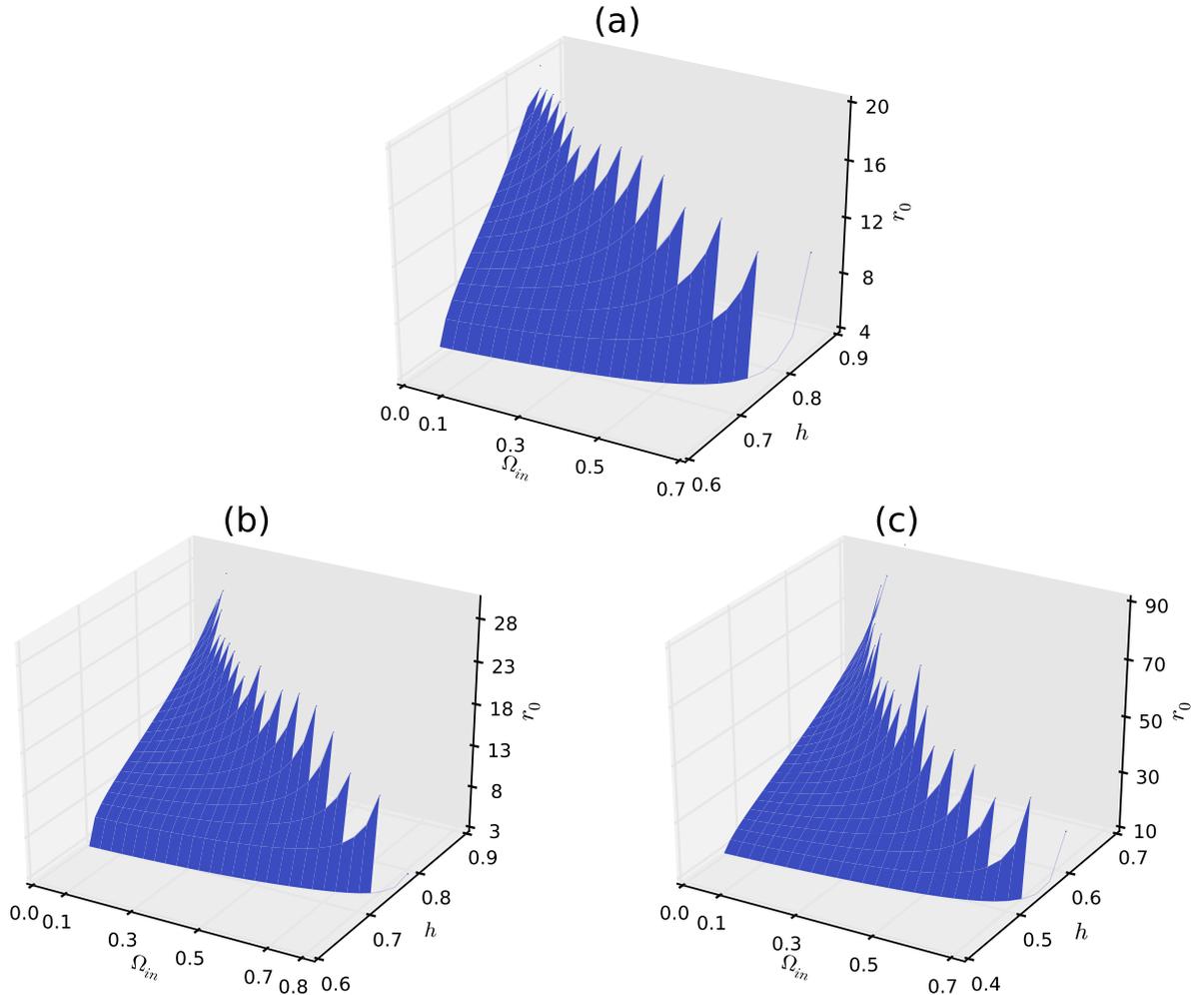}
 \caption{\label{fig:gauss3d}
 The 3D plot of the allowed parameter space of the Gaussian
 model for (a) OHRO at $z=1.43$, (b) OHRO at $z=1.55$, (c)
 OHRO at $z=3.91$, respectively. The blue contours indicate
 the model parameters making the theoretical cosmic age
 equal to the age of OHRO at the same redshift. The allowed
 parameter spaces are the upper regions of these contours.
 Note that $r_0$ is in units of Gpc. See the text for details.}
 \end{figure}
 \end{center}

%======================================================================

\vspace{-10mm} % used here just for a more comfortable typesetting

%============================= Fig. Gauss 2d =================================

 \begin{center}
 \begin{figure}[b]
 \centering
 \includegraphics[width=0.7\textwidth]{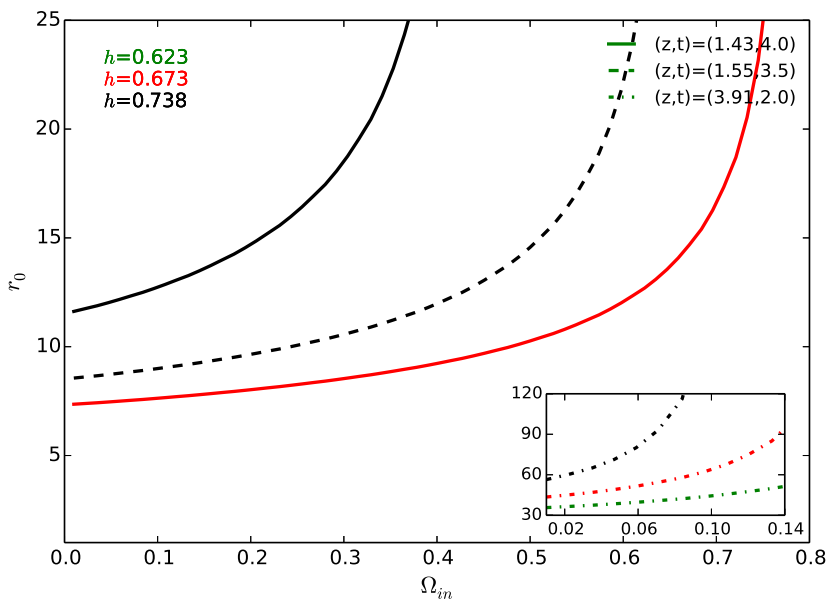}
 \caption{\label{fig:gauss2d}
 The 2D plot of the allowed parameter space of the Gaussian
 model with fixed $h=0.738$ (black contour lines), $0.673$
 (red contour lines), $0.623$ (green contour lines) for OHROs
 at $z=1.43$ (solid contour lines), $z=1.55$ (dashed contour
 lines), $z=3.91$ (dash-dotted contour lines), respectively.
 The contour lines indicate the model parameters making the
 theoretical cosmic age equal to the age of OHRO at the same
 redshift. The allowed parameter spaces are the upper regions
 of these contour lines. Note that $r_0$ is in units of Gpc.
 See the text for details.}
 \end{figure}
 \end{center}

%======================================================================

\vspace{-10mm} % used here just for a more comfortable typesetting

%============================= Fig. Gauss ps =================================

 \begin{center}
 \begin{figure}[t]
 \centering
 \includegraphics[width=0.68\textwidth]{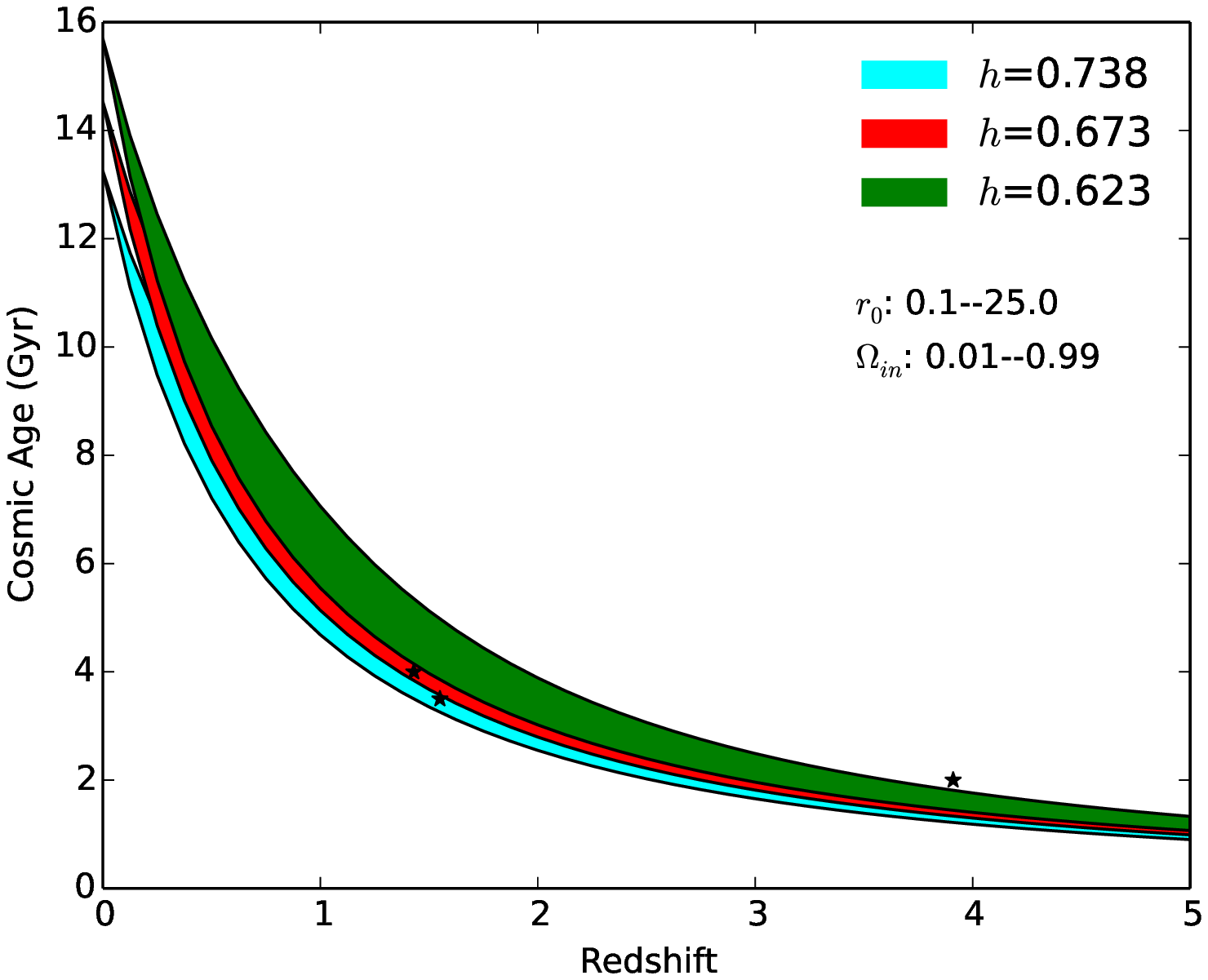}
 \caption{\label{fig:gaussps}
 Cosmic age as function of redshift $z$ for the parameter
 space $0.01\leq\Omega_{in}\leq 0.99$, $0.1\,{\rm Gpc}\leq
 r_0\leq 25\,{\rm Gpc}$ of the Gaussian model with fixed
 $h=0.738$ (cyan), $0.673$ (red), $0.623$ (green). The three
 OHROs at redshift $z=1.43$, $1.55$, $3.91$ are also indicated
 by black stars. See the text for details.}
 \end{figure}
 \end{center}

%======================================================================

\vspace{-10mm} % used here just for a more comfortable typesetting

%============================= section 3.1 ===================================

\subsection{The Gaussian model}\label{sec3a}

The gradient in the bang time $t_B(r)$ corresponds to a
 currently non-vanishing decaying mode \cite{Zibin:2008vj,
 Silk:1977xx}, which might imply an inhomogeneous early
 universe that violates inflation, and lead to inhomogeneities
 in the galaxy formation time. To be simple, one might assume
 that the big bang is spatially homogeneous, namely $t_B$ is
 a constant. Following e.g. \cite{Wang:2011kj,GarciaBellido:2008nz},
 we can set $t_B=0$ for convenience. In this
 case, Eq.~(\ref{eq19}) becomes
 \be{eq25}
 H_{\perp 0} (r)=H_0\mathcal{F}(\Omega_M)\,,
 \ee
 where the function $\cal F$ is given in Eq.~(\ref{eq20}), and
 \be{eq26}
 H_0\equiv 1/t_0\,.
 \ee
 So, in this case, one only needs to specify $\Omega_M (r)$,
 and then $H_{\perp 0} (r)$ can be found from Eq.~(\ref{eq25}).

At first, we consider the simplest Gaussian LTB void model
 \cite{Wang:2011kj,Clifton:2008hv}, in which the matter
 density function $\Omega_M(r)$ has a Gaussian profile, namely
 \be{eq27}
 \Omega_M(r)=1+\left(\Omega_{in}-1\right)
 \,\exp\left(-\frac{r^2}{2r_0^2}\right)\,,
 \ee
 where $\Omega_{in}$ is the matter
 density at the center of the void, and $r_0$ describes the
 size of the void. In this work, we only consider the case of
 $\Omega_{in}<1$. From Eq.~(\ref{eq25}), it is easy to obtain
 \be{eq28}
 H_{\perp 0}(r) = H_0\frac{\sqrt{\Omega_K(r)} -
 \Omega_M(r)\sinh^{-1}\sqrt{\frac{\Omega_K(r)}{\Omega_M(r)}}}
 {\left[\Omega_K(r)\right]^{3/2}}\,,
 \ee
 where $\Omega_K(r)=1-\Omega_M(r)$, and $H_0$ actually plays
 the role of Hubble constant. So, there are three free model
 parameters, namely $\Omega_{in}$, $r_0$, and $h$ (which is
 the Hubble constant $H_0$ in units of 100\,km/s/Mpc).

To test the Gaussian model with the three OHROs at redshift
 $z=1.43$, $1.55$, $3.91$, we scan a fairly wide
 parameter space $0.01\leq\Omega_{in}\leq 0.99$,
 $1.0\,{\rm Gpc}\leq r_0\leq 501.0\,{\rm Gpc}$,
 and $0.4\leq h\leq 1.0$. At every point, we numerically
 calculate the theoretical cosmic age at redshift $z=1.43$,
 $1.55$, $3.91$ for the Gaussian model with the corresponding
 parameters $\Omega_{in}$, $r_0$, and $h$. Then, we obtain
 three contours which indicate the model parameters making
 the theoretical cosmic age equal to the age of OHRO at the
 same redshift $z=1.43$, $1.55$, $3.91$. We present them in
 Fig.~\ref{fig:gauss3d}. Only the parameters corresponding
 to a theoretical cosmic age larger than (or equal to) the
 age of OHRO at the same redshift are allowed. In fact, the
 allowed parameter spaces are the upper regions of the
 contours shown in Fig.~\ref{fig:gauss3d}. From
 Fig.~\ref{fig:gauss3d}, it is easy to see that a large $r_0$
 is required to accommodate the three OHROs. In particular,
 from the panel~(c) of Fig.~\ref{fig:gauss3d}, we find that
 $r_0>10\,{\rm Gpc}$ is required to accommodate OHRO at
 redshift $z=3.91$. To see this clearer, in
 Fig.~\ref{fig:gauss2d} we show the 2D slices of the allowed
 parameter space with fixed $h=0.738$, $0.673$, $0.623$. Note
 that $h=0.738$ is the best-fit value of the Hubble constant
 from SHOES SNIa project \cite{Riess:2011yx}; $h=0.673$ is the
 one from Planck CMB data \cite{Ade:2013zuv}. On the other
 hand, Sandage {\it et al.} advocated a lower Hubble constant
 from HST SNIa, and the best-fit value of their final result is
 $h=0.623$ \cite{Sandage:2006cv}. The allowed parameter spaces
 are the upper regions of the contour lines in
 Fig.~\ref{fig:gauss2d}. Note that the absence of green-solid
 contour line in Fig.~\ref{fig:gauss2d} means that the entire
 plotted parameter space of the Gaussian model with a fixed
 $h=0.623$ is allowed for OHRO at $z=1.43$. This fact can be
 seen clearly from Fig.~\ref{fig:gaussps}, in which we scan the
 parameter space $0.01\leq\Omega_{in}\leq 0.99$,
 $0.1\,{\rm Gpc}\leq r_0\leq 25\,{\rm Gpc}$ with fixed
 $h=0.738$, $0.673$, $0.623$, and plot cosmic age as function
 of redshift $z$. It is clear that OHRO at $z=1.43$ is below
 the lower boundary of the green region, and hence all the
 parameter space is allowed in this case. Similarly, the
 absence of red-dashed and green-dashed contour lines in
 Fig.~\ref{fig:gauss2d} means that the entire plotted parameter
 space of the Gaussian model with fixed $h=0.673$, $0.623$ is
 allowed for OHRO at $z=1.55$, and it can also be seen clearly
 from Fig.~\ref{fig:gaussps}, since OHRO at $z=1.55$ is below
 the lower boundaries of both the red and green regions. On the
 other hand, from Fig.~\ref{fig:gaussps}, one can also see that
 at least $r_0>25\,{\rm Gpc}$ is required to accommodate OHRO
 at $z=3.91$, since it is above all the upper boundaries of the
 cyan, red, and green regions. In fact, from the small panel
 in Fig.~\ref{fig:gauss2d}, at least $r_0>30\,{\rm Gpc}$ is
 required to accommodate OHRO at $z=3.91$. The unusually large
 $r_0$ brings a serious crisis to the Gaussian LTB void model.
 As is well known, the Hubble radius (Hubble horizon)
 $H_0^{-1}\simeq 3.0\,h^{-1}\,{\rm Gpc}$~\cite{Kolb:1990xx}
 characterizes the size of our observable universe. The size
 of the void should be much larger than the size of our
 observable universe to accommodate OHROs (see the discussion
 in Sec.~\ref{sec5} however). On the other hand,
 there is a serious tension between this unusually large
 $r_0$ and the much lower $r_0$ of order 1.0\,Gpc inferred from
 other observations (e.g. SNIa, CMB and so on) mentioned in
 Sec.~\ref{sec1}. If the Gaussian LTB void model can be
 consistent with other observations (e.g. SNIa, CMB and so
 on), it cannot accommodate OHROs. Of course, it is
 known that the lower Hubble constant, the larger cosmic age
 is. However, as shown in the panel~(c) of
 Fig.~\ref{fig:gauss3d}, $r_0>10\,{\rm Gpc}$ is still required
 even for a very low $h=0.4$. So, this serious crisis cannot
 be alleviated with a lower Hubble constant.

%============================= Fig. CGBH 3d =================================

 \begin{center}
 \begin{figure}[tb]
 \centering
 \includegraphics[width=1.0\textwidth]{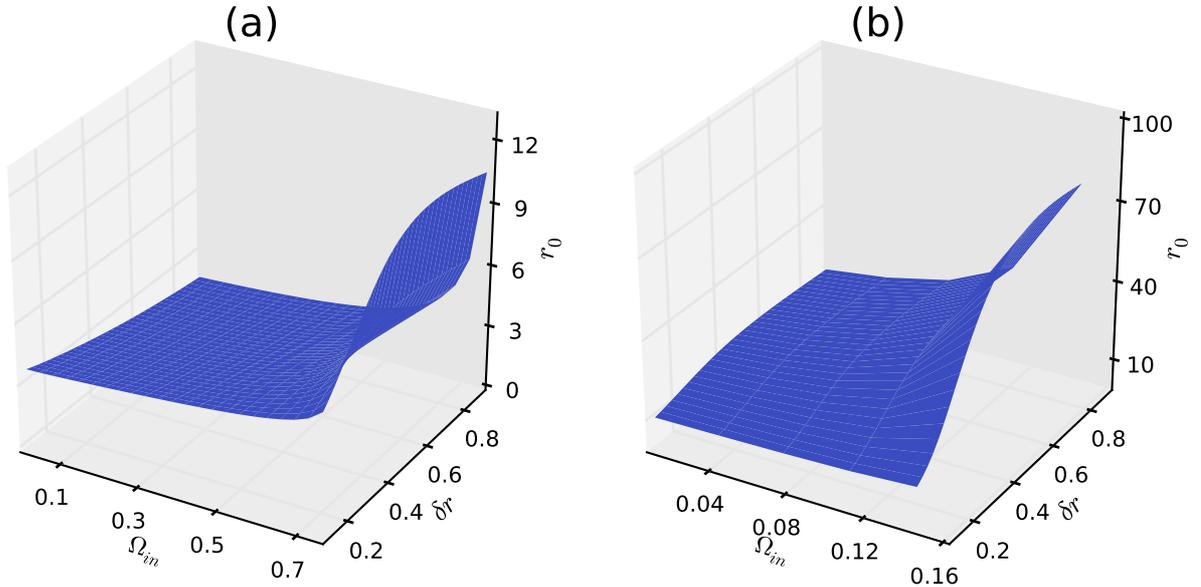}
 \caption{\label{fig:cgbh3d}
 The same as in Fig.~\ref{fig:gauss3d}, except for the CGBH
 model with a fixed $h=0.673$, and (a) OHRO at $z=1.43$,
 (b) OHRO at $z=3.91$. See the text for details.}
 \end{figure}
 \end{center}

%======================================================================

\vspace{-12mm} % used here just for a more comfortable typesetting

%============================= Fig. CGBH ps =================================

 \begin{center}
 \begin{figure}[tb]
 \centering
 \includegraphics[width=0.68\textwidth]{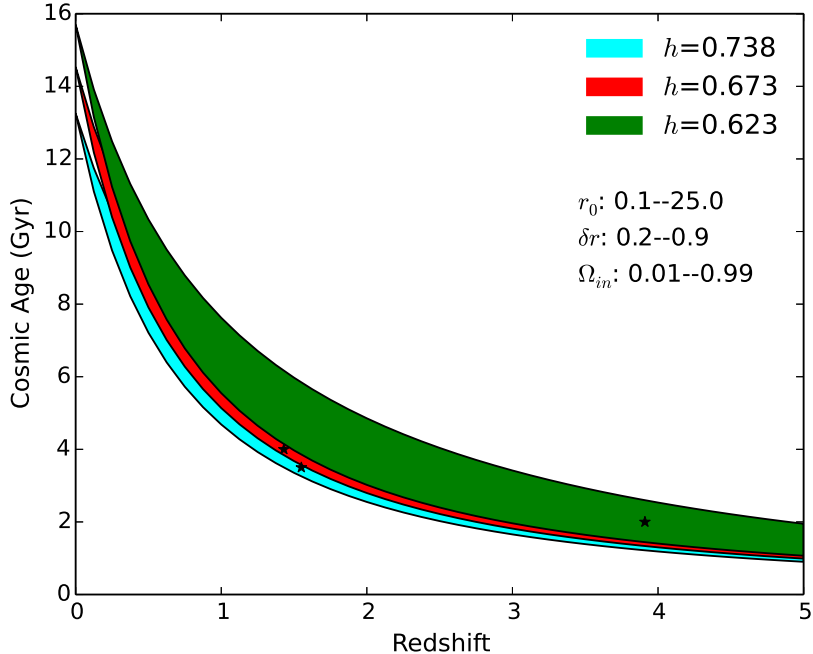}
 \caption{\label{fig:cgbhps}
 The same as in Fig.~\ref{fig:gaussps}, except for the CGBH
 model with an additional parameter
 $0.2\leq\delta r\leq 0.9$. See the text for details.}
 \end{figure}
 \end{center}

%======================================================================

\vspace{-15mm} % used here just for a more comfortable typesetting

%============================= section 3.2 ===================================

\subsection{The CGBH model}\label{sec3b}

Next, we consider a simplified version of the so-called
 Garcia-Bellido-Haugb{\o}lle (GBH) model \cite{GarciaBellido:2008nz},
 namely the constrained GBH (CGBH) model \cite{GarciaBellido:2008nz}
 (see also e.g. \cite{Wang:2011kj}). In CGBH model, one also assumes
 that the big bang is spatially homogeneous, namely $t_B$ is
 a constant which can be set to zero. So, Eq.~(\ref{eq25}) is
 valid in the CGBH model. The matter density function $\Omega_M(r)$
 is given by \cite{GarciaBellido:2008nz} (see also e.g.
 \cite{Wang:2011kj})
 \be{eq29}
 \Omega_M(r)=1+\left(\Omega_{in}-1\right)
 \left\{\frac{1-\tanh[(r-r_0)/2\Delta r]}
 {1+\tanh(r_0/2\Delta r)}\right\}\,,
 \ee
 where $\Omega_{in}$ is the matter density at the center of
 the void; $r_0$ describes the size of the void; $\Delta r$
 characterizes the transition to uniformity. In this work, we
 only consider the case of $\Omega_{in}<1$.
 From Eq.~(\ref{eq25}), we get
 \be{eq30}
 H_{\perp 0}(r) = H_0\frac{\sqrt{\Omega_K(r)} -
 \Omega_M(r)\sinh^{-1}\sqrt{\frac{\Omega_K(r)}{\Omega_M(r)}}}
 {\left[\Omega_K(r)\right]^{3/2}}\,,
 \ee
 where $\Omega_K(r)=1-\Omega_M(r)$, and $H_0$ actually plays
 the role of Hubble constant. So, there are four free model
 parameters, namely $\Omega_{in}$, $r_0$, $h$ (which is
 the Hubble constant $H_0$ in units of 100\,km/s/Mpc), and
 $\delta r\equiv \Delta r/r_0$ (which is equivalent to
 $\Delta r$ in fact).

\begin{center}
 \begin{figure}[tb]
 \centering
 \includegraphics[width=1.0\textwidth]{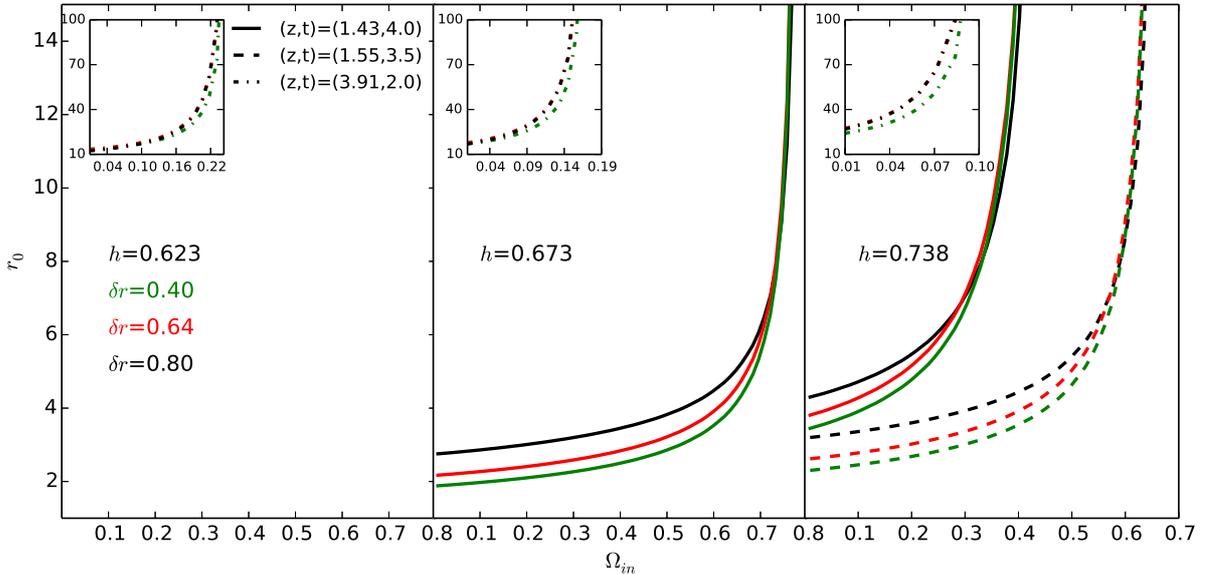}
 \caption{\label{fig:cgbh2d}
 The same as in Fig.~\ref{fig:gauss2d}, except for the CGBH
 model with fixed $h=0.623$ (left panel), $0.673$ (middle
 panel), $0.738$ (right panel), and $\delta r=0.40$ (green
 contour lines), $0.64$ (red contour lines), $0.80$ (black
 contour lines), for OHROs at $z=1.43$ (solid contour lines),
 $z=1.55$ (dashed contour lines), $z=3.91$ (dash-dotted
 contour lines). See the text for details.}
 \end{figure}
 \end{center}

Similar to the previous subsection, we firstly scan the full
 parameter space to test this model with OHROs. However, since
 there are four free parameters in the CGBH model, it is
 difficult to plot a 4D parameter space. Instead, we consider
 the 3D plot of the allowed parameter space of the CGBH model
 with a fixed $h=0.673$ coming from Planck CMB
 data \cite{Ade:2013zuv}, and we present it in
 Fig.~\ref{fig:cgbh3d}. Note that the wide parameter ranges
 we scanned are $0.01\leq\Omega_{in}\leq 0.99$,
 $1.0\,{\rm Gpc}\leq r_0\leq 501.0\,{\rm Gpc}$,
 and $0.1\leq\delta r\leq 0.9$. The absence of plot for OHRO
 at redshift $z=1.55$ in Fig.~\ref{fig:cgbh3d} means that the
 entire plotted parameter space of the CGBH model with a fixed
 $h=0.673$ is allowed for OHRO at $z=1.55$. This can be seen
 clearly from Fig.~\ref{fig:cgbhps} in which OHRO at $z=1.55$
 is below the lower boundary of the red region, and hence the
 entire parameter space is allowed in this case. Note that from
 the panel~(b) in Fig.~\ref{fig:cgbh3d}, a large
 $r_0>10\,{\rm Gpc}$ is required to accommodate OHRO at
 $z=3.91$. Also, in Fig.~\ref{fig:cgbh2d} we show the 2D
 slices of the allowed parameter space with fixed $h=0.623$,
 $0.673$, $0.738$, and $\delta r=0.40$, $0.64$, $0.80$. Note
 that $\delta r=0.64$ is the best-fit value from SNIa, CMB
 and BAO \cite{GarciaBellido:2008nz}, and $\delta r=0.40$ and
 $0.80$ are close to the edges of its $2\sigma$ confidence
 region. The absence of the contour lines for OHROs at $z=1.43$
 and $1.55$ in the left panel of Fig.~\ref{fig:cgbh2d} means
 that the entire plotted parameter space of the CGBH model with
 a fixed $h=0.623$ is allowed for these two OHROs. And the
 absence of the contour lines for OHRO at $z=1.55$ in the
 middle panel of Fig.~\ref{fig:cgbh2d} means that the entire
 plotted parameter space of the CGBH model with a fixed
 $h=0.673$ is allowed for OHRO at $z=1.55$. This can be seen
 clearly from Fig.~\ref{fig:cgbhps} in which OHRO at $z=1.43$
 is below the lower boundary of the green region, and OHRO at
 $z=1.55$ is below the lower boundaries of both the red
 and green regions. From the three small panels
 of Fig.~\ref{fig:cgbh2d} and the panel~(b) of
 Fig.~\ref{fig:cgbh3d}, we see that at least
 $r_0>10\,{\rm Gpc}$ is required to accommodate OHRO at
 $z=3.91$. Therefore, the same crisis also exists in the
 CGBH model. Again, the size of the void should be much
 larger than the size of our observable universe (characterized
 by the Hubble radius/horizon
 $H_0^{-1}\simeq 3.0\,h^{-1}\,{\rm Gpc}$~\cite{Kolb:1990xx})
 to accommodate OHROs (see the discussion in Sec.~\ref{sec5}
 however). On the other hand, there is a serious tension
 between this unusually large $r_0>10\,{\rm Gpc}$ and the much
 lower $r_0$ of order 1.0\,Gpc inferred from other observations
 (e.g. SNIa, CMB and so on) mentioned in Sec.~\ref{sec1}.
 If the CGBH LTB void model can be consistent with other
 observations (e.g. SNIa, CMB and so on), it cannot accommodate
 OHROs.

%============================= Fig. CGBH 2d =================================

%======================================================================

\vspace{-10mm} % used here just for a more comfortable typesetting

%============================= section 3.3 ===================================

\subsection{The GBH model}\label{sec3c}

Finally, we consider the original version of GBH model
 \cite{GarciaBellido:2008nz}, in which one does not assume
 that the big bang is spatially homogeneous. Therefore,
 Eqs.~(\ref{eq25}) and (\ref{eq26}) are invalid, and hence
 $\Omega_M(r)$ and $H_{\perp 0}(r)$ should be specified
 independently. In GBH model, they are given by
 \cite{GarciaBellido:2008nz}
 \bea
 \Omega_M(r)=\Omega_{out}+\left(\Omega_{in}-\Omega_{out}\right)
 \left\{\frac{1-\tanh[(r-r_0)/2\Delta r]}
 {1+\tanh(r_0/2\Delta r)}\right\}\,,\label{eq31}\\[2mm]
 H_{\perp 0}(r)=H_{out}+\left(H_{in}-H_{out}\right)\left\{
 \frac{1-\tanh[(r-r_0)/2\Delta r]}
 {1+\tanh(r_0/2\Delta r)}\right\}\,,\label{eq32}
 \eea
 where $\Omega_{out}$ is the asymptotic value of the matter
 density; $\Omega_{in}$ is the matter density at the center
 of the void; $H_{out}$ and $H_{in}$ describe the Hubble
 expansion rate outside and inside the void, respectively;
 $r_0$ describes the size of the void; $\Delta r$
 characterizes the transition to uniformity. Following
 \cite{GarciaBellido:2008nz}, we fix $\Omega_{out}=1$. So,
 there are five free model parameters, namely $\Omega_{in}$,
 $r_0$, $\delta r\equiv \Delta r/r_0$ (which is equivalent
 to $\Delta r$ in fact), $h_{in}$ and $h_{out}$ (which
 are $H_{in}$ and $H_{out}$ in units of 100\,km/s/Mpc).

\begin{center}
 \begin{figure}[bht]
 \centering
 \includegraphics[width=1.0\textwidth]{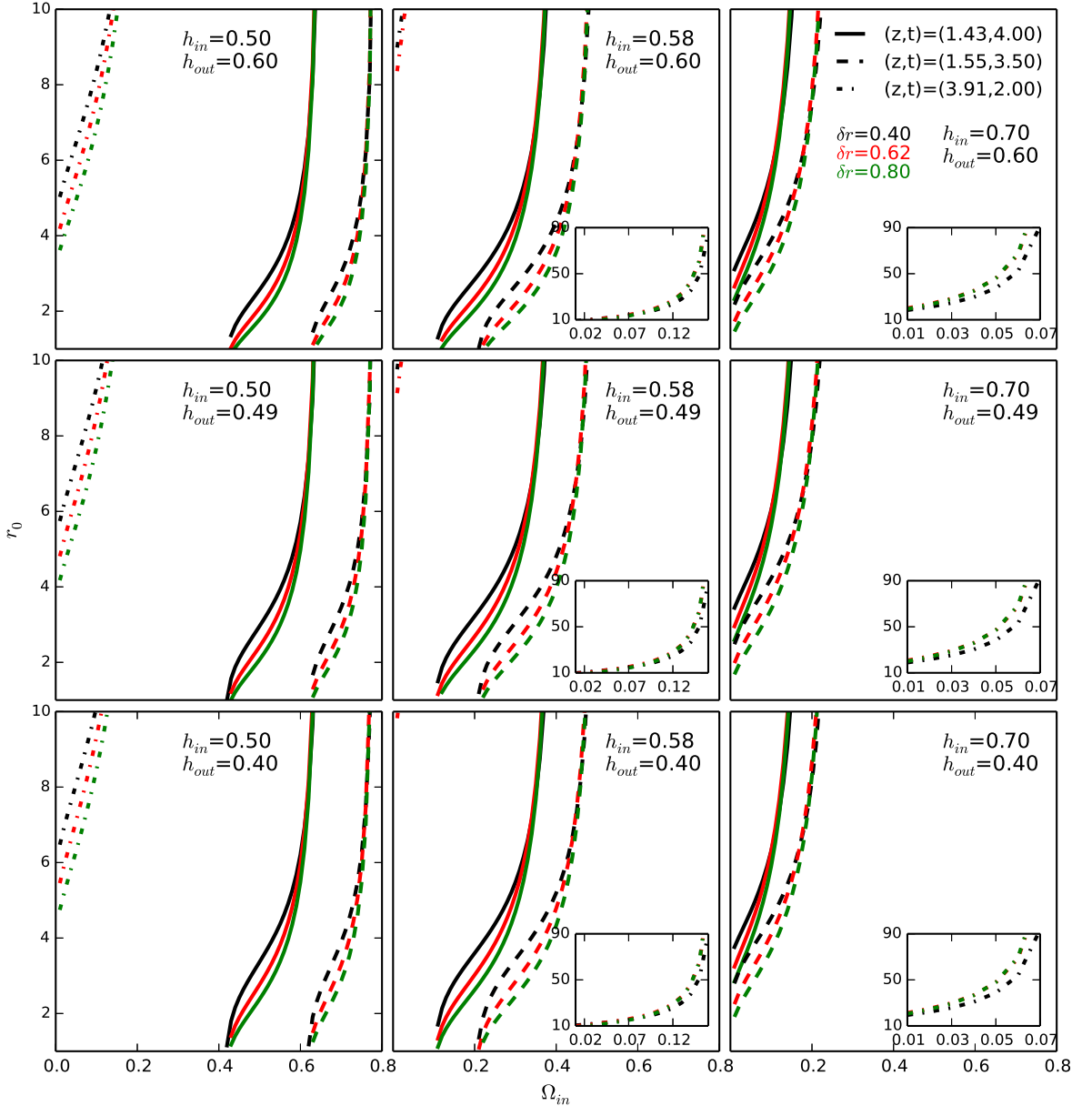}
 \caption{\label{fig:gbh2d}
 The same as in Fig.~\ref{fig:gauss2d}, except for the GBH
 model with fixed $h_{in}=0.50$ (left column), $0.58$ (middle
 column), $0.70$ (right column), $h_{out}=0.60$ (top row),
 $0.49$ (middle row), $0.40$ (bottom row), $\delta r=0.40$
 (black contour lines), $0.62$ (red contour lines), $0.80$
 (green contour lines), for OHROs at $z=1.43$ (solid contour
 lines), $z=1.55$ (dashed contour lines), $z=3.91$ (dash-dotted
 contour lines). See the text for details.}
 \end{figure}
 \end{center}

Similar to the previous subsections, we try to scan the full
 parameter space to test this model with OHROs. However, since
 there are five free parameters in the GBH model, it is very
 difficult to plot a 5D parameter space. Instead,
 in Fig.~\ref{fig:gbh2d} we show the 2D slices of the allowed
 parameter space with fixed $h_{in}=0.50$, $0.58$, $0.70$, and
 $h_{out}=0.60$, $0.49$, $0.40$, as well as $\delta r=0.40$,
 $0.62$, $0.80$. Note that $h_{in}=0.58$, $h_{out}=0.49$,
 $\delta r=0.62$ are the best-fit values from SNIa, CMB and
 BAO \cite{GarciaBellido:2008nz}, and we appropriately vary
 these parameters to see their effect on the allowed parameter
 space. From Fig.~\ref{fig:gbh2d}, it is easy to see that the
 parameters $\delta r$ and $h_{out}$ have fairly minor effects
 on the allowed parameter space. On the other hand, comparing
 the three columns of Fig.~\ref{fig:gbh2d}, we find that the
 parameter $h_{in}$ plays a considerable role. The smaller
 $h_{in}$, the wider parameter space can be allowed. From the
 middle and right columns of Fig.~\ref{fig:gbh2d}
 (especially from the small panels), it is easy to see that for
 $h_{in}\,\gsim\,0.58$, a large $r_0>10\,{\rm Gpc}$ is required
 to accommodate OHRO at $z=3.91$. In this case, as in the
 previous two LTB void models, the serious crisis also exists
 in the GBH LTB void model. That is, the size of the void
 should be much larger than the size of our observable universe
 to accommodate OHROs (see the discussion in Sec.~\ref{sec5}
 however); there exists a serious tension between this
 unusually large $r_0>10\,{\rm Gpc}$ and the much lower $r_0$
 of order 1.0\,Gpc inferred from other observations (e.g. SNIa,
 CMB and so on) mentioned in Sec.~\ref{sec1}. However, for a
 very low $h_{in}=0.50$, the required $r_0$ can be in a lower
 range $\sim 4-6\,{\rm Gpc}$ to accommodate OHROs, as shown in
 the left column of Fig.~\ref{fig:gbh2d}. Note that the Hubble
 radius/horizon $H_0^{-1}\simeq 3.0\,h^{-1}\,{\rm Gpc}\sim
 6\,{\rm Gpc}$ for a very low $h\sim 0.50$. So, in this case,
 it is possible to accommodate OHROs while the size of the
 void is smaller than the size of our observable universe.
 However, if we further consider the constraints from other
 observations (e.g. SNIa, CMB and so on), the situation
 becomes subtle. In \cite{GarciaBellido:2008nz}, the best-fit
 parameters with $2\sigma$ uncertainties from SNIa, CMB and
 BAO are given by $h_{in}=0.58\pm 0.03$, $h_{out}=0.49\pm 0.2$,
 $\Omega_{in}=0.13\pm 0.06$, $r_0=2.3\pm 0.9\,{\rm Gpc}$,
 $\delta r=0.62\pm(>0.20)$. There exists still a remarkable
 tension far beyond $2\sigma$ between OHROs and other observations,
 because $h_{in}=0.50$ and $r_0\sim 4-6\,{\rm Gpc}$ can be
 excluded by other observations (e.g. SNIa, CMB and BAO) far beyond
 $2\sigma$ regions. Even worse, the effect of $h_{in}$ is in
 contrast to the one of $r_0$ actually. If we increase $h_{in}$, the
 required $r_0$ to accommodate OHRO at $z=3.91$ will increase
 correspondingly, as shown in Fig.~\ref{fig:gbh2d}. Therefore,
 it is very difficult to conciliate both $h_{in}$ and $r_0$ with the
 higher $h_{in}=0.58\pm 0.03$ and the smaller $r_0=2.3\pm 0.9$
 inferred from SNIa, CMB and BAO \cite{GarciaBellido:2008nz} at the
 same time. We are in a dilemma. The tension between the constraints
 from OHROs and other observations (e.g. SNIa, CMB and so on)
 is fairly serious. So, the age problem cannot be completely
 alleviated in the GBH LTB void model, although it is in a situation
 slightly better than the Gaussian model and the CGBH model (but at
 the price of having more model parameters).

%============================= Fig. GBH 2d =================================

%======================================================================

\vspace{-15mm} % used here just for a more comfortable typesetting

%============================= Fig. Gauss relaxed =================================

 \begin{center}
 \begin{figure}[tbh]
 \centering
 \includegraphics[width=1.0\textwidth]{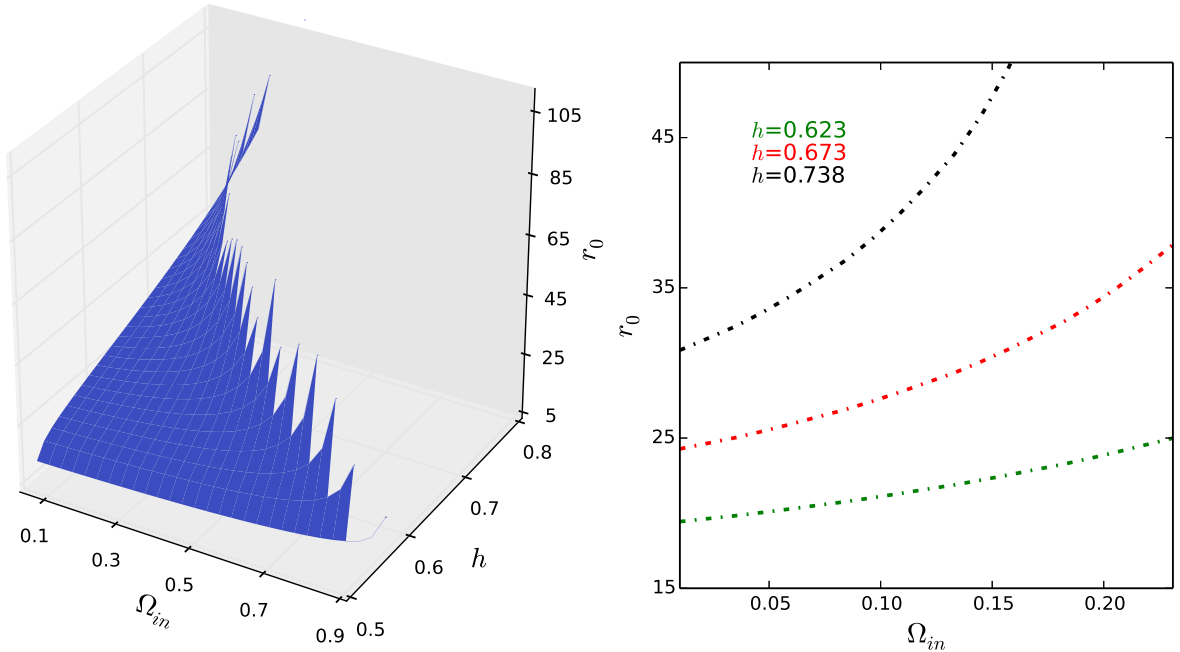}
 \caption{\label{fig:gauss17}
 The same as in Fig.~\ref{fig:gauss3d} (left panel) and
 Fig.~\ref{fig:gauss2d} (right panel), except only for OHRO
 at $z=3.91$ whose age has been changed to 1.7\,Gyr. See the
 text for details.}
 \end{figure}
 \end{center}

%======================================================================

\vspace{-10mm} % used here just for a more comfortable typesetting

%============================= section 4 ===================================

\section{Alleviating the age problem}\label{sec4}

In the previous section, it is easy to find that the age
 problem in three LTB void models is mainly due to the OHRO at
 $z=3.91$. In fact, this OHRO has ruled out (at least brought
 trouble to) most cosmological models (see
 e.g.~\cite{Alcaniz:1999kr,Friaca:2005ba,darkenergytested:01,
 darkenergytested:02,Yang:2009ae,Wei:2007ig,Zhang:2007ps,Wei:2010hs,
 Wang:2010su,Capozziello:2007gr,Movahed:2007cs,Movahed:2007ie,
 Movahed:2007ps,Pires:2006rd}), including the well-known
 concordance $\Lambda$CDM model. Naturally, one might doubt
 on the validity of this quasar APM 08279+5255 at redshift $z=3.91$
 (we thank the referee for pointing out this issue). In fact, its
 age was estimated model-dependently. In~\cite{Komossa:2002cn},
 the age 2.0\,--\,3.0\,Gyr was obtained by using the giant
 elliptical model (M4a) and the extreme model (M6a), which are
 two of 12 chemical evolution models considered
 in~\cite{Hamann:1993iu}. In~\cite{Yang:2009ae}, its age was
 re-evaluated following~\cite{Hamann:1993iu}. They found the
 age of APM 08279+5255 {\em since the initial star formation
 and stellar evolution in the galaxy}: (1)~the best estimated
 value is 2.1\,Gyr; (2) $1\sigma$ lower limit is 1.8\,Gyr;
 (3) the lowest limit is 1.5\,Gyr (although this is highly
 improbable as noted in~\cite{Yang:2009ae}). However, this is
 not the age since the beginning of the universe, because the
 initial star formation started about 0.2\,--\,0.3\,Gyr after
 the big bang. Therefore, in~\cite{Yang:2009ae} they concluded
 the age of APM 08279+5255 {\em since the beginning of the
 universe}: (1) the best estimated value is 2.3\,Gyr; (2)
 $1\sigma$ lower limit is 2.0\,Gyr; (3) the lowest limit is
 1.7\,Gyr (although this is highly improbable as noted
 in~\cite{Yang:2009ae}).

In the previous section, we used the age 2.0\,Gyr for APM
 08279+5255 at $z=3.91$, which is just the $1\sigma$ lower
 limit given in~\cite{Yang:2009ae}. Here, we would like to
 consider the lowest limit 1.7\,Gyr~\cite{Yang:2009ae}, and
 see whether the age problem can be alleviated. Because the
 age problem is most serious in the Gaussian LTB void model
 as mentioned above, and not to break the length limit, we
 only consider the Gaussian model here. In
 Fig.~\ref{fig:gauss17}, we show the 3D plot and the 2D slices
 of the allowed parameter space of the Gaussian model only for
 OHRO at $z=3.91$ whose age has been changed to 1.7\,Gyr.
 Although $r_0\,\gsim\,20\,{\rm Gpc}$ is still required for
 $h>0.623$ (see the right panel of Fig.~\ref{fig:gauss17}), we
 find from the left panel of Fig.~\ref{fig:gauss17} that $r_0$
 can reach $\sim 5\,{\rm Gpc}$ for a lower $h\,\lsim\,0.55$.
 In this case, $r_0\sim 5\,{\rm Gpc}$ to accommodate OHROs
 can be slightly smaller than the Hubble radius/horizon
 $H_0^{-1}\simeq 3.0\,h^{-1}\,{\rm Gpc}$~\cite{Kolb:1990xx}.
 However, $r_0\sim 5\,{\rm Gpc}$ is still larger than the
 one of order 1.0\,Gpc inferred from other observations (e.g.
 SNIa, CMB and so on) mentioned in Sec.~\ref{sec1}, the tension
 still exists. Nevertheless, comparing with the previous
 section, it is fair to say that the tension has been greatly
 soften and the age problem is alleviated in some sense.

\begin{center}
 \begin{figure}[tb]
 \centering
 \vspace{-2.1mm} % used here just for a more comfortable typesetting
 \includegraphics[width=0.66\textwidth]{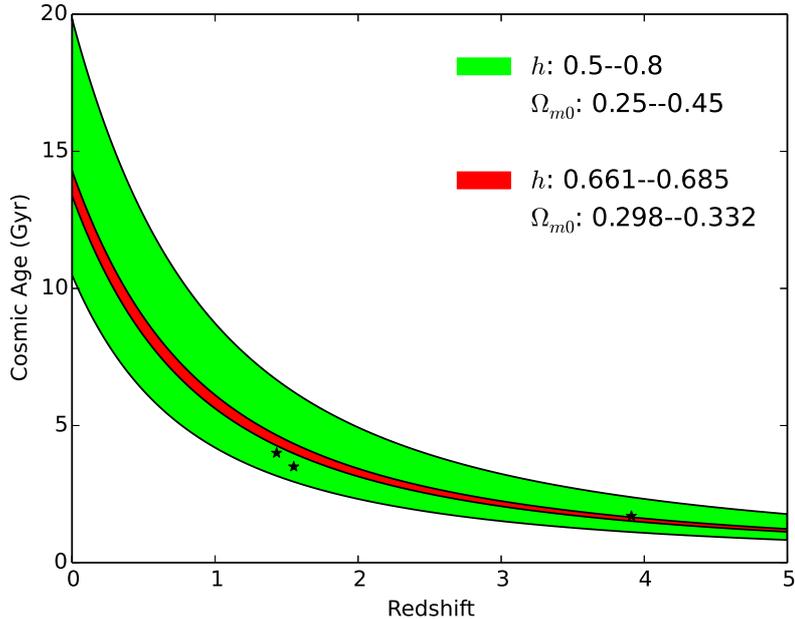}
 \caption{\label{fig:lcdmps}
 Cosmic age as function of redshift $z$ for the parameter
 space (a) $0.5\leq h\leq 0.8$, $0.25\leq\Omega_{m0}\leq 0.45$
 (green) and (b) $0.661\leq h\leq 0.685$,
 $0.298\leq\Omega_{m0}\leq 0.332$ (red) of the
 flat $\Lambda$CDM model. The three OHROs at redshift $z=1.43$,
 $1.55$, $3.91$ are also indicated by black stars. Note that
 the age of OHRO at $z=3.91$ has been changed to 1.7\,Gyr. See
 the text for details.}
 \end{figure}
 \end{center}

Since the age 2.0\,Gyr for APM 08279+5255 at $z=3.91$ has also
 ruled out most cosmological models including $\Lambda$CDM
 model, it is of interest to see whether the age problem can
 also be alleviated in $\Lambda$CDM model, if the age of
 APM 08279+5255 is changed to the lowest limit
 1.7\,Gyr~\cite{Yang:2009ae}. For the flat $\Lambda$CDM model,
 its age at redshift $z$ is given
 by~\cite{Alcaniz:1999kr,darkenergytested:01,Wei:2007ig,Kolb:1990xx}
 \be{eq33}
 T(z)=\int_z^\infty \frac{d\tilde{z}}{(1+\tilde{z})H(\tilde{z})}\,,
 ~~{\rm where}~~H(z)=H_0\sqrt{\Omega_{m0}(1+z)^3+(1-\Omega_{m0})}\,.
 \ee
 There are two free parameters, namely $\Omega_{m0}$ and $h$
 (the Hubble constant $H_0$ in units of 100\,km/s/Mpc). In
 Fig.~\ref{fig:lcdmps}, we scan the parameter space (a)
 $0.5\leq h\leq 0.8$, $0.25\leq\Omega_{m0}\leq 0.45$ and
 (b) $0.661\leq h\leq 0.685$, $0.298\leq\Omega_{m0}\leq 0.332$,
 and plot cosmic age as function of redshift $z$. Note that
 the latter (b) is in fact the $1\sigma$ region from Planck
 CMB data~\cite{Ade:2013zuv}, namely $h=0.673\pm 0.012$,
 $\Omega_{m0}=0.315\pm 0.017$. From Fig.~\ref{fig:lcdmps}, we
 see that all the three OHROs can be well accommodated in
 $\Lambda$CDM model for the wide parameter space~(a), since the
 cosmic age can be larger than the ones of all OHROs. Even for
 the narrow parameter space~(b), OHRO at $z=3.91$ is just on
 the edge. Clearly, the age problem can also be alleviated in
 $\Lambda$CDM model if the age of APM 08279+5255 is changed
 to the lowest limit 1.7\,Gyr.

%============================= Fig. LCDM ps =================================

%======================================================================

\vspace{-11mm} % used here just for a more comfortable typesetting

%============================= section 5 ===================================

\section{Conclusion and discussion}\label{sec5}

As is well known, one can explain the current cosmic acceleration by
 considering an inhomogeneous and/or anisotropic universe
 (which violates the cosmological principle), without invoking dark
 energy or modified gravity. The well-known one of this kind of
 models is the so-called Lema\^{\i}tre-Tolman-Bondi (LTB) void
 model, in which the universe is spherically symmetric and radially
 inhomogeneous, and we are living in a locally underdense void
 centered nearby our location. In the present work, we test various
 LTB void models with some old high redshift objects (OHROs).
 Obviously, the universe cannot be younger than its constituents. We
 find that an unusually large $r_0$ (characterizing the size of the
 void) is required to accommodate these OHROs in LTB void models.
 There is a serious tension between this unusually large $r_0$
 and the much smaller $r_0$ inferred from other observations
 (e.g. SNIa, CMB and so on). However, if we instead consider
 the lowest limit 1.7\,Gyr for the quasar APM 08279+5255 at
 redshift $z=3.91$, this tension could be greatly alleviated.

It is worth noting that in addition to the three OHROs used in
 this work, there are other OHROs in the literature, for
 instance, the 4.0\,Gyr old radio galaxy 3C~65 at $z=1.175$
 \cite{Stockton:1995xx}, and the high redshift quasar B1422+231
 at $z=3.62$ whose best-fit age is 1.5\,Gyr with a lower bound
 of 1.3\,Gyr~\cite{Yoshii:1998bw}. However, they cannot be used
 to constrain the models as restrictive as the three OHROs used
 in this work. So, we do not consider them here. On the other
 hand, 9 extremely old globular clusters in M31 galaxy
 \cite{Ma:2009wb,Wang:2010qn} were considered in
 \cite{Wang:2010su}. Note that their ages are estimated to be
 in the range $14-16\,{\rm Gyr}$ \cite{Ma:2009wb,Wang:2010qn},
 which is much larger than the total age of the universe
 $\sim 13.8\,{\rm Gyr}$ inferred from the CMB observations
 (e.g.~WMAP~\cite{Hinshaw:2012aka}
 and Planck~\cite{Ade:2013zuv}). Of course, this does not
 mean that they cannot be used in the relevant works. However,
 since as mentioned in Sec.~\ref{sec1} we only use OHROs in
 the present work, and hence we also have not considered these
 9 extremely old globular clusters in M31 galaxy.

In Sec.~\ref{sec3}, we find that an unusually
 large $r_0>10\,{\rm Gpc}$ (or even larger) is required to
 accommodate OHROs, which means that the size of the void
 should be much larger than the size of our
 observable universe (characterized by the Hubble
 radius/horizon
 $H_0^{-1}\simeq 3.0\,h^{-1}\,{\rm Gpc}$~\cite{Kolb:1990xx}).
 However, this does not make the LTB void models invalid (we
 thank the referee for pointing out this issue). Instead, it
 just means that the whole void is unobservable, or likewise
 that the void is a super-horizon mode perturbation. But
 since the variation of density inside the horizon is not
 negligible, such a model is physically distinct from FRW
 model, and hence is meaningful in principle (we thank the
 referee for pointing out this issue). In the present work,
 the age problem manifests itself mainly in the serious
 tension between the constraints from OHROs and
 other observations (e.g. SNIa, CMB and so on).

%============================= acknowledgements ===================================

\section*{ACKNOWLEDGEMENTS}
We thank the anonymous referee for quite useful comments
 and suggestions, which helped us to improve this work.
 We are grateful to Profs. Rong-Gen~Cai, Shuang~Nan~Zhang
 and Tong-Jie~Zhang for helpful discussions. We also thank
 Hao~Wang, Si-Qi~Liu, as well as Zu-Cheng~Chen, Jing~Liu,
 Ya-Nan~Zhou, Xiao-Bo~Zou and Hong-Yu~Li for kind help and
 discussions. This work was supported in part by NSFC under
 Grants No.~11175016 and No.~10905005, as well as NCET
 under Grant No.~NCET-11-0790.

\renewcommand{\baselinestretch}{1.0}

%============================= references ==================================

\end{document}